\documentclass[journal=jctcce,manuscript=article,layout=traditional]{achemso}

\usepackage{geometry, times, amsmath, amsfonts, indentfirst, pdfpages, graphicx, caption, subcaption, booktabs, array, multirow, hyperref, appendix, mathtools, pdflscape, ulem}

\author{Letif Mones}
\email{lam81@cam.ac.uk}
\affiliation{Mathematics Institute, University of Warwick, Coventry CV4 7AL, United Kingdom}
\alsoaffiliation{Engineering Laboratory, University of Cambridge, Trumpington Street, Cambridge, CB2 1PZ, United Kingdom}

\author{G\'abor Cs\'anyi}
\affiliation{Engineering Laboratory, University of Cambridge, Trumpington Street, Cambridge, CB2 1PZ, United Kingdom}

\author{Christoph Ortner}
\affiliation{Mathematics Institute, University of Warwick, Coventry CV4 7AL, United Kingdom}

\title{Preconditioners for the geometry optimisation and saddle point search of molecular systems}

\graphicspath{{./}{../figures/}}

\date{}

\begin{document}

\maketitle

\section{Abstract}

A  class of preconditioners is introduced to enhance geometry optimisation and transition state search of molecular systems. We start from the Hessian of molecular mechanical terms, decompose it and retain only its positive definite part to construct a sparse preconditioner matrix. The construction requires only the computation of the gradient of the corresponding molecular mechanical terms that are already  available in popular force field software  packages.
For molecular crystals, the preconditioner can be combined straightforwardly with the exponential preconditioner recently introduced for periodic systems. The efficiency is demonstrated on several systems using empirical, semiempirical and  ab initio potential energy surfaces.

\section{Introduction}

Geometry optimisation and transition state search are fundamental procedures to identify important stationary points of molecules, molecular crystals and material systems in computational chemistry. Since the evaluation of chemically accurate ab initio potential energies and  gradients are computationally demanding, several techniques have been developed over the last three decades to enhance the convergence of optimisation methods.

Among  the most widely used  are quasi-Newton methods, in particular
BFGS or its limited memory version \cite{Liu89}, which start with a the identity as its guess for the Hessian and update it at each iteration based on the gradient information collected from previous steps.
Initializing with  Hessian guess that include more molecule specific geometrical information can improve the convergence. For instance, just introducing some connectivity information about a molecule can lead to surprisingly good results\cite{Vebjorn02,Packwood16}.

More sophisticated methods estimate the initial  Hessian guess from a surrogate potential (e.g. force field or semiempirical potential) whose second derivative can be obtained at low computational cost. The update of the  approximate Hessian  can then be achieved either by using quasi-Newton methods or other techniques such as DIIS \cite{Csaszar84, Vogel93}. Such strategies significantly improve the speed of convergence in either Cartesian or internal coordinates.\cite{Baker93}

If the model Hessian  is cheap   to calculate (e.g., as obtained from a surrogate
model) and provides a reasonable approximation of the quantum Hessian, it may be advantageous to recompute it at every optimisation step. Such a scheme was introduced by Lindh et al. \cite{Lindh95}, where a model potential is constructed consisting of quadratic terms for all distances, angles and dihedrals in the molecule. At each geometry optimisation step the  force field is constructed such that  the current conformation is its local minimum, and its  ``Hessian'' is computed, neglecting the dependence of the force field parameters on the geometry. With this construction, the model ``Hessian'' is therefore not the Hessian of any potential. Nevertheless, this approach yields excellent performance, which led to its wide implementation~\cite{Werner12}.

The model ``Hessian'' of Lindh \cite{Lindh95} can be considered as a preconditioner or metric that effects a transformation to a new coordinate system where the optimisation problem is better conditioned, hence  algorithms converge more rapidly and tend to be more robust. Geometrically, the shape of the energy landscape becomes more isotropic. In general it is desirable that both the construction and inversion of the preconditioner matrix are inexpensive (at least compared to the computation of  the energy and gradient). This can be achieved by building a sparse  preconditioner from simple analytical functions.
Since the preconditioner defines a metric in  configuration space it needs to be positive definite. This requirement is automatically fulfilled in the Lindh approach\cite{Lindh95} by the use of quadratic molecular mechanical terms with equilibrium points corresponding to the actual geometry at each step.

We recently introduced a general and effective preconditioner for geometry optimisation and saddle search for material systems \cite{Packwood16}. This preconditioner is determined by the local connectivity structure of atoms making both the construction and inversion computationally inexpensive. Especially for larger systems we observe an order of magnitude or larger reduction of the number of optimisation steps for the preconditioned LBFGS method compared to the  same without preconditioning.
Here we expand our previous work to molecules and molecular crystals by combining it with a force field based preconditioner inspired by the approach of Lindh et al. \cite{Lindh95}.

\section{Methods}

\subsection{Enhancing geometry optimisation by using preconditioners}

We briefly review the methodology for preconditioning geometry optimisation and the dimer saddle search method for material systems\cite{Packwood16}. For a system with $N$ particles let $x_{k} \in \mathbb{R}^{3N}$ denote the configuration at the $k$th iterate of an optimisation algorithm. The corresponding energy, gradient and preconditioner are denoted by $f_{k}=f(x_{k})$, $g_{k} = \nabla f (x_{k})$ and $P_k \in \mathbb{R}^{3N \times 3N} \approx \nabla^2 f(x_k)$, respectively. A preconditioned steepest-descent step is then given by
\begin{equation}
        \label{eq:quasi_newton}
        x_{k+1} = x_{k} - \alpha_{k}P_{k}^{-1}g_{k}
\end{equation}
where $\alpha_{k}$ is the step size obtained from some line search procedure at the $k$th iteration. If $P_k = I$, then \eqref{eq:quasi_newton} becomes the standard steepest descent scheme and if $P_k = \nabla^2 f(x_k)$, then it becomes a quadratically convergent Newton scheme. In general, different choices of $P_k$ may ``interpolate'' between these extremes.

From an alternative point of view preconditioning can be thought of as a coordinate transformation, where a new set of coordinates is defined as $y_{k} := P^{1/2}x_{k}$.
The advantage of this framework is that once an appropriate preconditioner matrix is available then any optimisation algorithm can be modified by applying the original algorithm on the transformed coordinates. To obtain the final form of the modified algorithm we need to transform  the variables back to the original coordinate system.
As a simple example, applying the coordinate transformation on the gradient descent equation immediately leads to the equation of quasi Newton schemes (eq. \ref{eq:quasi_newton}):
\begin{align}
        \label{eq:example}
        \begin{split}
                y_{k} &= y_{k-1} - \alpha_{k} \nabla_{y} F(y_{k}) \\
                P^{1/2}_{k} x_{k} & = P^{1/2}_{k} x_{k-1} - \alpha_{k} P^{-1/2}_{k} \nabla_{x} f(x_{k}) \\
                x_{k} & = x_{k-1} - \alpha_{k} P^{-1}_{k} \nabla_{x} f(x_{k}) \\
        \end{split}
\end{align}
where we used the fact that $\nabla_y F(y_{k}) = P^{-1/2}_{k} \nabla_{x} f(x_{k})$.
Preconditioning popular optimisation methods like LBFGS, conjugate gradients, FIRE\cite{Bitzek06} etc. is similarly possible.

A simple preconditioner that is effective for a wide range of materials systems is based on the following $N \times N$ matrix \cite{Packwood16}
\begin{equation}
        \label{eq:exp_precond_L}
        L_{ij} =
        \begin{cases}
                -\mu \exp \left( -A \left( \frac{r_{ij}}{r_{\mathrm{nn}}} - 1\right) \right), & i \ne j \text{~and~} |r_{ij}| < r_{\mathrm{cut}} \\
                0, & i \ne j \text{~and~} |r_{ij}| \ge r_{\mathrm{cut}} \\
                -\sum_{i \ne j} L_{ij}, & i=j
        \end{cases}
\end{equation}
where $i$ and $j$ denote atomic indices and $\mu$, $A$, $r_{\mathrm{cut}}$ and $r_{\mathrm{nn}}$ are parameters that can be user-specified or estimated numerically.
We note that \eqref{eq:exp_precond_L} is a generalisation of the Laplacian matrix used to represent undirected graphs. Given a specific connectivity defined by $r_{\mathrm{cut}}$ and setting $A = 0$ and $\mu = 1$, $L_{ij}$ reduces exactly to the Laplacian matrix.
The actual $3N \times 3N$ preconditioner is simply obtained from the corresponding $L_{ij}$ element in an isotropic manner:
\begin{equation}
        \label{eq:exp_precond}
        [P_\mathrm{Exp}]_{i+3(k-1), j+3(l-1)} =
        \begin{cases}
                L_{ij}, & k=l \\
                0, & k \ne l
        \end{cases}
\end{equation}
where the $k$ and $l$ indices denote Cartesian components.
Despite its simplicity in capturing only geometric connectivity  but no specific material information, $P_{\rm Exp}$ provides a good model for the local curvature of the potential energy landscape, which effectively controls ill-conditioning in  large systems.  The application of $P_{\rm Exp}$ resulted in a significant reduction of the number of optimisation steps required  for several material systems compared to the unpreconditioned LBFGS\cite{Packwood16}.

\subsection{FF-based preconditioners}

Preliminary tests showed that for molecular systems such as molecules in gas phase or molecular crystals using $P_{\rm Exp}$ still leads to a speed-up, but a much more modest one than for material systems. The explanation for this is that molecular systems contain a wide range of different interactions (e.g., pair, angle, dihedral, electrostatic, dispersive)  of vastly varying stiffness which in addition are more loosely coupled, and this creates a second source of ill-conditioning distinct from ill-conditioning due to large system size. Inspired by the use of internal coordinates in molecular optimisation techniques\cite{Fogarasi92} and the model Hessian of Lindh et al. \cite{Lindh95} we therefore propose a generalisation of $P_{\rm Exp}$ that is effective  also for molecular systems.

The construction of our \textit{FF-based preconditioner} begins with a surrogate potential energy function, given by a sum over internal coordinates each describing a short-range bond in the system (distance, angle, or dihedral),
\begin{equation}
        \label{eq:Vtot}
        V_{\mathrm{FF}} = \sum \limits_{\alpha} V_{\alpha}(\xi_{\alpha}(\mathbf{r})).
\end{equation}
The individual potential energy terms are in general simple functions of the internal coordinates. Some examples of most typical forms are the quadratic, Morse or torsional potentials, respectively given by
\begin{align}
        \label{eq:V_quadratic}
        V_{\mathrm{Quadratic}}(q) &= \frac{1}{2} k (q - q_{0})^{2},  \\
        \label{eq:bond_morse}
        V_{\mathrm{Morse}}(d) &= D_{0}(1 - \exp(-\alpha (d - d_{0})))^{2}, \quad \text{and}  \\
        \label{eq:V_torsion}
        V_{\mathrm{Torsion}}(\phi) &= \frac{1}{2} k_{\phi} (1 + \cos(n \phi - \phi_{0}))
\end{align}
where the corresponding parameters can be taken from standard force field libraries.

Due to its simple functional form $H_{\rm FF} := \nabla^2 V_{\mathrm{FF}}$ is cheap to compute. If only short-range bonds are taken into account then it is also sparse, hence it is cheap to store and invert. Moreover, we expect that $V_{\rm FF}$ gives a good qualitative approximation to the quantum potential energy landscape, hence $H_{\rm FF}$ is a good qualitative approximation to the quantum Hessian $\nabla^2 f$. Therefore, $H_{\rm FF}$ satisfies all the conditions required for a preconditioner \textit{except} that it will in general be indefinite. A conceptually straightforward but computationally expensive approach to overcome this limitation is to enforce positivity by replacing all eigenvalues of $H_{\rm FF}$ with their absolute values. Instead, we propose to analytically modify the local Hessian contributions to ensure overall positivity, resulting  in a further \textit{reduction} in computational cost.

The Hessian contribution from $V_\alpha$ is given by
\begin{equation}
        \label{eq:hessian}
        \mathbf{H}_{\alpha} = \frac{\partial^{2} V_{\alpha}}{\partial \mathbf{r}^{2}} = \underbrace{\frac{\partial \xi_{\alpha}}{\partial \mathbf{r}} \otimes \frac{\partial \xi_{\alpha}}{\partial \mathbf{r}} \frac{\partial^{2} V_{\alpha}}{\partial \xi_{\alpha}^{2}}}_{\mathbf{H}_{\alpha}^{(1)}} + \underbrace{\frac{\partial^{2} \xi_{\alpha}}{\partial \mathbf{r}^{2}} \frac{\partial V_{\alpha}}{\partial \xi_{\alpha}}}_{\mathbf{H}_{\alpha}^{(2)}},
\end{equation}
where we have decomposed $H_\alpha$ into two terms, $\mathbf{H}_{\alpha}^{(1)}$ and $\mathbf{H}_{\alpha}^{(2)}$. If $V_\alpha$ is quadratic then $\frac{\partial^{2} V_{\mathrm{Quadratic}}}{\partial \xi^{2}} = k > 0$, hence $\textbf{H}_\alpha^{(1)}$ is positive semi-definite, while the sign of $\textbf{H}_\alpha^{(2)}$
is ambiguous. Note however, that if the system is at equilibrium of $V_\alpha$ with respect to $\xi_{\alpha}$, i.e., if
$\frac{\partial V_{\alpha}}{\partial \xi_{\alpha}} = 0$, then $\textbf{H}_\alpha^{(2)} = 0$. This in fact is the case in the Lindh approach~\cite{Lindh95}.
Instead of adjusting $V_\alpha$ at every step such that the geometry corresponds to its equilibrium, here we simply drop $\textbf{H}_\alpha^{(2)}$ and only use $\mathbf{H}_{\alpha}^{(1)}$ to construct the preconditioner, thus ensuring that it always stays positive definite.

For non-quadratic contributions we expect that $\frac{\partial^{2} V_{\alpha}}{\partial \xi_{\alpha}^{2}} > 0$ for most but not all bonds, hence we enforce positivity by replacing it with its absolute value. This leads to the following general preconditioner for molecular systems:
\begin{equation}
        \label{eq:ff_precond}
        P_{\mathrm{FF}} = \sum \limits_{\alpha} \tilde{\mathbf{H}}_{\alpha}^{(1)} = \sum \limits_{\alpha} \frac{\partial \xi_{\alpha}}{\partial \mathbf{r}} \otimes \frac{\partial \xi_{\alpha}}{\partial \mathbf{r}} \left| \frac{\partial^{2} V_{\alpha}}{\partial \xi_{\alpha}^{2}} \right|
\end{equation}
It is worth noting that $\frac{\partial \xi_{\alpha}}{\partial \mathbf{r}}$ is already computed by molecular mechanics force field based MD programs since it is required for the assembly of $\nabla V_\alpha$. Thus, the only new quantity that must be computed is $\frac{\partial^{2} V_{\alpha}}{\partial \xi_{\alpha}^{2}}$, which represents a negligible additional computational cost.

We note that the final functional form \eqref{eq:ff_precond} of our preconditioner is very similar to that of Lindh et al. \cite{Lindh95}, however we arrived at it from a fundamentally different perspective, which has several advantages. Lindh et al.'s method was introduced for quadratic terms only, hence the force constants have to be recomputed after each optimisation step (as the equilibrium bond lengths, angles and dihedrals are set to the actual ones to obtain positive semidefinite matrix). Our method can be considered as a generalisation of their approach, allowing arbitrary functional forms of internal coordinate dependent terms. In particular this means that the FF parameters need not be adjusted to achieve a positive semidefinite matrix, and incorporating different parameter sets is straightforward. Moreover, as we will see in \S~\ref{sec:FF+Exp} and in \S~\ref{sec:materials} our perspective makes it easy to extend the preconditioner construction to new situations.

\subsection{Combining FF and Exp preconditioners}
\label{sec:FF+Exp}

For molecular crystal,  intermolecular interactions  also play an important role. This consideration led us to combine the molecular mechanics based FF preconditioner (describing bonded interactions) with the Exp preconditioner \eqref{eq:exp_precond} (tuned to describe non-bonded interactions):
\begin{equation}
        \label{eq:exp_ff_precond}
        P_{\mathrm{Exp+FF}} = P_{\mathrm{Exp}} + P_{\mathrm{FF}}
\end{equation}
$P_{\rm FF}$ is fully specified from the chosen force field $V_{\rm FF}$.
To specify $P_{\rm Exp}$ we first manually choose the parameters $A$ and $r_{\mathrm{cut}}$ in \eqref{eq:exp_precond} to account for the interaction between molecules\cite{Packwood16}. The remaining parameters are computed in a similar automatic manner as described in Ref.~\cite{Packwood16}, and we keep only those matrix elements of $P_{\rm Exp}$ for which the corresponding matrix element in $P_{\rm FF}$ is zero.
We note that correct scaling between $P_{\rm FF}$ and $P_{\rm Exp}$ is implicitly ensured via the $\mu$ parameter in Eq~\ref{eq:exp_precond_L}.

\subsection{Implementation details}

We tested the FF and FF+Exp preconditioners on a range of optimisation and saddle search tasks. For geometry optimisations the form of the preconditioned LBFGS method was identical to the one we describe in ref\cite{Packwood16}: At each iterate the search direction is given by
\begin{align}
        \label{eq:lbfgs}
        \begin{split}
                &{\bf input}~q = \nabla f(x_{k}) \\
                &s_{k} = x_{k} - x_{k-1} \\
                &y_{k} = \nabla f(x_{k}) - \nabla f(x_{k-1}) \\
                &\rho_{k} = 1/y_{k}^{T} s_{k} \\
                &\mathrm{for } \: i = k,\ldots,k-m \\
                &\qquad \alpha_i = \rho_i s_i^T q \\
                &\qquad q = q - \alpha_i y_i \\
                &\fbox{$z = P_k^{-1} q$} \\
                &\mathrm{for } \: i = k-m,\ldots,k \\
                &\qquad \beta_i = \rho_i y_i^T z\\
                &\qquad z = z + (\alpha_i - \beta_i)s_i \\
                &{\bf output}~p_{k} = z \\
        \end{split}
\end{align}
with initial search direction $z = P_{0}^{-1} \nabla f(x_{0})$. The step length selection is obtained by a backtracking line-search enforcing only the Armijo condition\cite{Packwood16}.

For saddle search tasks we slightly modified the superlinearly converging dimer method \cite{Kastner08}. Dimer methods use two copies of the system with coordinates $x_{1}$ and $x_{2}$ and a fixed separation length $l = |x_{1}-x_{k}|$ between them.
The algorithm is usually split into two alternating steps \cite{Henkelman99}: (1) in the rotation step we fix the midpoint and rotate the endpoints to approximately align them with the lowest (negative) eigenmode of the Hessian; (2) in the translation step we shift the dimer to maximise the energy along the dimer direction while minimising energy in all directions perpendicular to it.

In principle, both the rotation and translation steps can be preconditioned, however, we found that in many systems preconditioning the rotation step results in a smaller spectral gap and hence slower convergence. Therefore, we chose to precondition only the translation step. Our implementation employs the conjugate gradient method using the Polak-Ribi\`ere formula:
\begin{align}
        \label{eq:cg}
        \begin{split}
                &{\bf input}~q_{k} = - (I - 2v_k \otimes v_k) \nabla f(x_{k}) \\
                &\fbox{$\beta = \frac{q_{k}^{T}P_{k}^{-1}(q_{k}-q_{k-1})}{q_{k-1}^{T}P_{k}^{-1}q_{k-1}}$} \\
                &s_{k} = q_{k} + \beta s_{k-1}\\
                &{\bf output}~\fbox{$p_{k} = \frac{P^{-1}s_{k}}{s_{k}^{T}P^{-1}s_{k}}$} \\
        \end{split}
\end{align}
and using the initial iterate $s_{0} = - (I - 2v_0 \otimes v_0)  \nabla f(x_{0})$. For computing the step length we used the trust region radius approach suggested by K\"astner and Sherwood\cite{Kastner08}, with acceptance criterion based on the projection of the gradient of the actual step.

For molecules in gas phase $V_{\rm FF}$ is invariant under rotations and translations, hence $P_{\rm FF}$ will be at least six fold degenerate with zero eigenvalues for any configuration of the molecule corresponding to the three translational and three rotational degrees of freedom. While these degrees of freedom could in principle be fixed we found that a straightforward solution is to simply regularise the preconditioner by replacing it with $P \leadsto P + c I$. We found that good generic values for $c$ are 0.1 and 1.0 eV / \AA$^{2}$ for, respectively, geometry optimisations and saddle search.

\subsection{Model systems and potentials}

\subsubsection{Organic molecules in gas phase}

Three potential energy surfaces were investigated for geometry optimisations: semiempirical PM6\cite{Stewart07}, DFT\cite{Hohenberg64, Kohn65} and MP2\cite{Moller34}. For the DFT potential we used the the BLYP exchange-correlation functional\cite{Becke88, Lee88} with the DZVP-MOLOPT basis set\cite{VandeVondele07} and plane wave cutoff of 480 Ry within the Gaussian and plane waves method (GPW) approach\cite{Lippert97} and Goedecker-Teter-Hutter (GTH) pseudopotentials\cite{Goedecker96}. For the MP2 calculations the 6-31G** basis set was applied.

In the case of geometry optimisations on the PM6 surface we compared three different force fields from which we constructed the preconditioners: the  force field of Lindh et al. (LFF)\cite{Lindh95}, the universal force field (UFF)\cite{Rappe92} and the generalised Amber force field (GAFF)\cite{Wang04}.

For transition state search we selected 7 examples from the benchmark of Baker and Chan\cite{Baker96} and three additional systems. The computations were performed on the semiempirical PM6 surface\cite{Stewart07}.

\subsubsection{Molecular crystals}

We compared four different optimisation schemes (unpreconditioned, only FF-based, only Exp-based and Exp+FF-based preconditioners) on five organic molecular crystals (systems XVIII to XXII) taken from the Organic Crystal Structure Prediction competition of the Cambridge Crystallographic Data Centre \cite{Bardwell11, Reilly16}. We used a DFT potential energy surface with the PBE exchange-correlation functional\cite{Perdew96} with a plane wave basis set using a cutoff energy of 800 eV and ultrasoft pseudopotentials\cite{Vanderbilt90}.

\subsubsection{Material systems}

We also tested how the force field based preconditioner works on two material systems compared to the exponential preconditioner. We examined the unpreconditioned and preconditioned geometry optimisation for bulk silicon and vacancy with varying system size. The potential energy surface was the screened Tersoff potential\cite{Tersoff86, Pastewka13} and we used the universal force field (UFF)\cite{Rappe92} for building the FF-based preconditioner matrix. Next we considered  bulk tungsten and a single interstitial site in bulk tungsten, also using different system sizes. The potential energy surface was a machine learning based Gaussian Approximation Potential (GAP) reproducing the quality of DFT (with PBE functional)~\cite{Szlachta14}. The preconditioner in this case was based on a simple Embedded Atom Method (EAM) potential~\cite{Daw84, Zhou04}.

\subsubsection{Software}

For the semiempirical PM6 method AmberTools16\cite{Case16, amber} was used. The MP2 potential surface was generated using MOLPRO\cite{Werner12, molpro}. The screened Tersoff potential was provided by Atomistica\cite{atomistica}. The DFT potentials with the BLYP and PBE functionals were, respectively, provided by CP2K\cite{VandeVondele05, cp2k} and CASTEP\cite{Segall02, castep}, using the QUIP interface\cite{Csanyi07, quip}. GAP model was  called via QUIP\cite{quip}.

In all cases the geometry optimisation was performed from within ASE\cite{Bahn02}. The other software packages were only used to compute the energy and gradient of the configurations. A Python implementation of the FF and Exp+FF preconditioners with several potential forms of nonbonded terms is available within ASE\cite{Bahn02, Larsen17} (https://gitlab.com/molet/ase).

\section{Results}

\subsubsection{Organic molecules in gas phase}

\begin{figure}
        \centering
        \includegraphics[width=1.0\columnwidth]{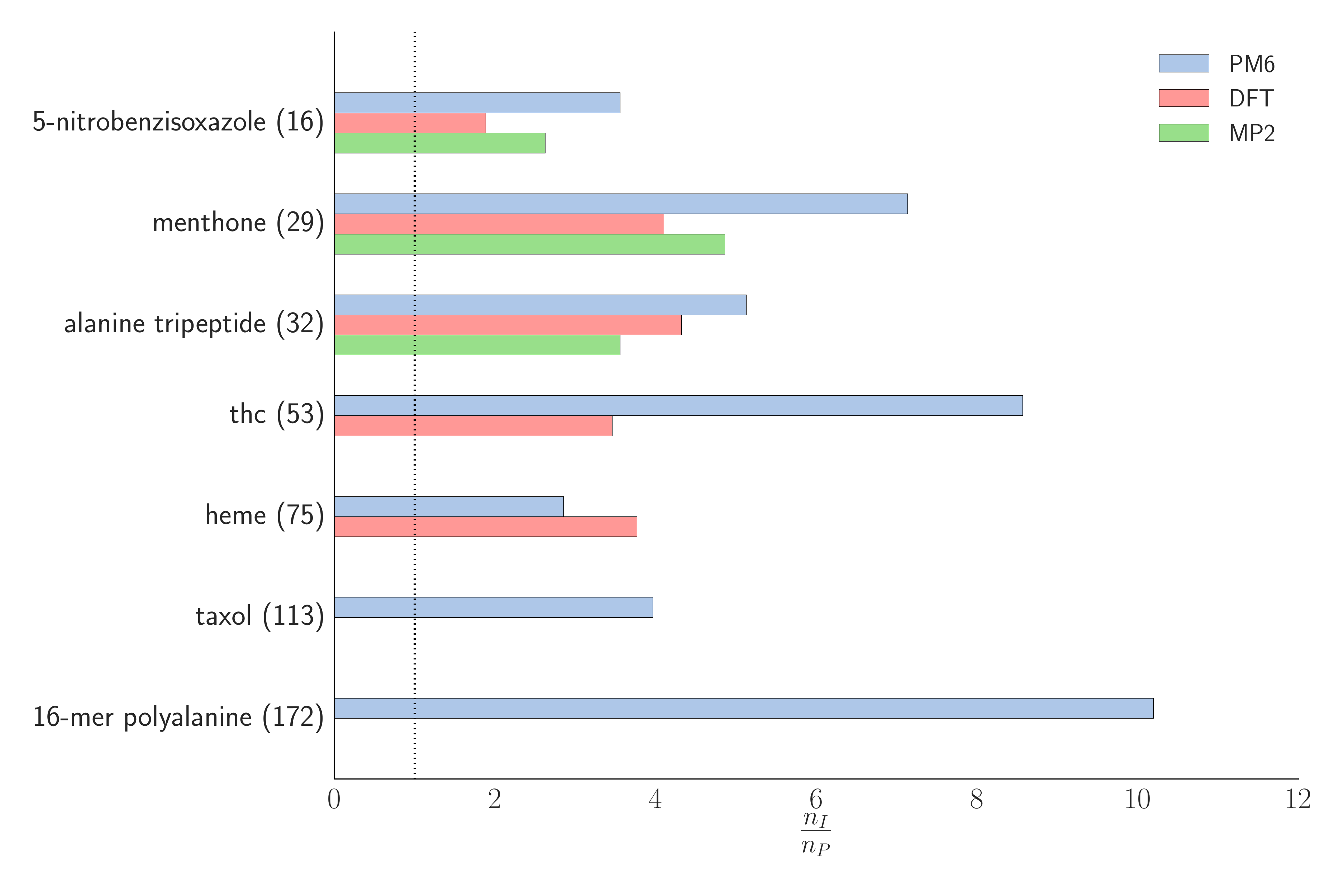}
        \caption{Computational saving of FF/GAFF preconditioner over unpreconditioned LBFGS optimisations for geometry optimisation of molecular systems in gas phase on three model potentials.}
        \label{fig:minim_mol_gas_id_ffgaff}
\end{figure}

We investigated several organic molecules' geometry optimisations with (FF) and without (ID) our new FF-based preconditioner on three potential energy surfaces (PM6, DFT and MP2), using the GAFF force field for building the preconditioner. The results are shown in Table~\ref{tab:minim_mol_gas_id_ffgaff} and Figure~\ref{fig:minim_mol_gas_id_ffgaff}. Convergence criterion of geometry optimisations was $|| \nabla E || = 10^{-3}$ eV \AA$^{-1}$ for DFT and MP2 surfaces while for the relatively inexpensive PM6 potential we applied a slightly tighter threshold of $|| \nabla E || = 10^{-4}$ eV \AA$^{-1}$. Depending on the system and underlying potential we can observe a 4--10 fold decrease in the required number of optimisation steps using our preconditioner.

\begin{table}
\centering
\begin{tabular}{llllllllll}
        \toprule
        \multirow{2}{*}{System (\# of atoms)} & \multicolumn{2}{c}{PM6} & \multicolumn{2}{c}{DFT(BLYP)} & \multicolumn{4}{c}{MP2/6-31G} \\
        & ID & FF/GAFF & ID & FF/GAFF & ID & FF/GAFF & FF/LFF & Lindh \\
        \midrule
        5-nitrobenzisoxazole (16) & 89 & 25 & 119 & 63 & 71 & 27 & 31 & 32 \\
        menthone (29) & 207 & 29 & 197 & 48 & 107 & 22 & 31 & 38 \\
        alanine tripeptide (32) & 395 & 77 & 536 & 124 & 210 & 59 & 47 & 67 \\
        thc (53) & 720 & 84 & 239 & 69 & & & & \\
        heme (75) & 500 & 175 & 358 & 95 & & & & \\
        taxol (113) & 1662 & 419 & & & & & & \\
        16-mer polyalanine (172) & 3549 & 348 & & & & & & \\
        \bottomrule
        \caption{Total number of function/gradient calls of geometry optimisation for organic molecules in gas phase using conventional (ID) and FF-based preconditioned (FF) LBFGS method on three different quantum chemistry surfaces. Convergence threshold was  $|| \nabla E || = 10^{-4}$ eV \AA$^{-1}$ for PM6 and $|| \nabla E || = 10^{-3}$ eV \AA$^{-1}$ for DFT and MP2 potentials, respectively.}
        \label{tab:minim_mol_gas_id_ffgaff}
\end{tabular}
\end{table}

\begin{figure}
        \centering
        \includegraphics[width=0.8\columnwidth]{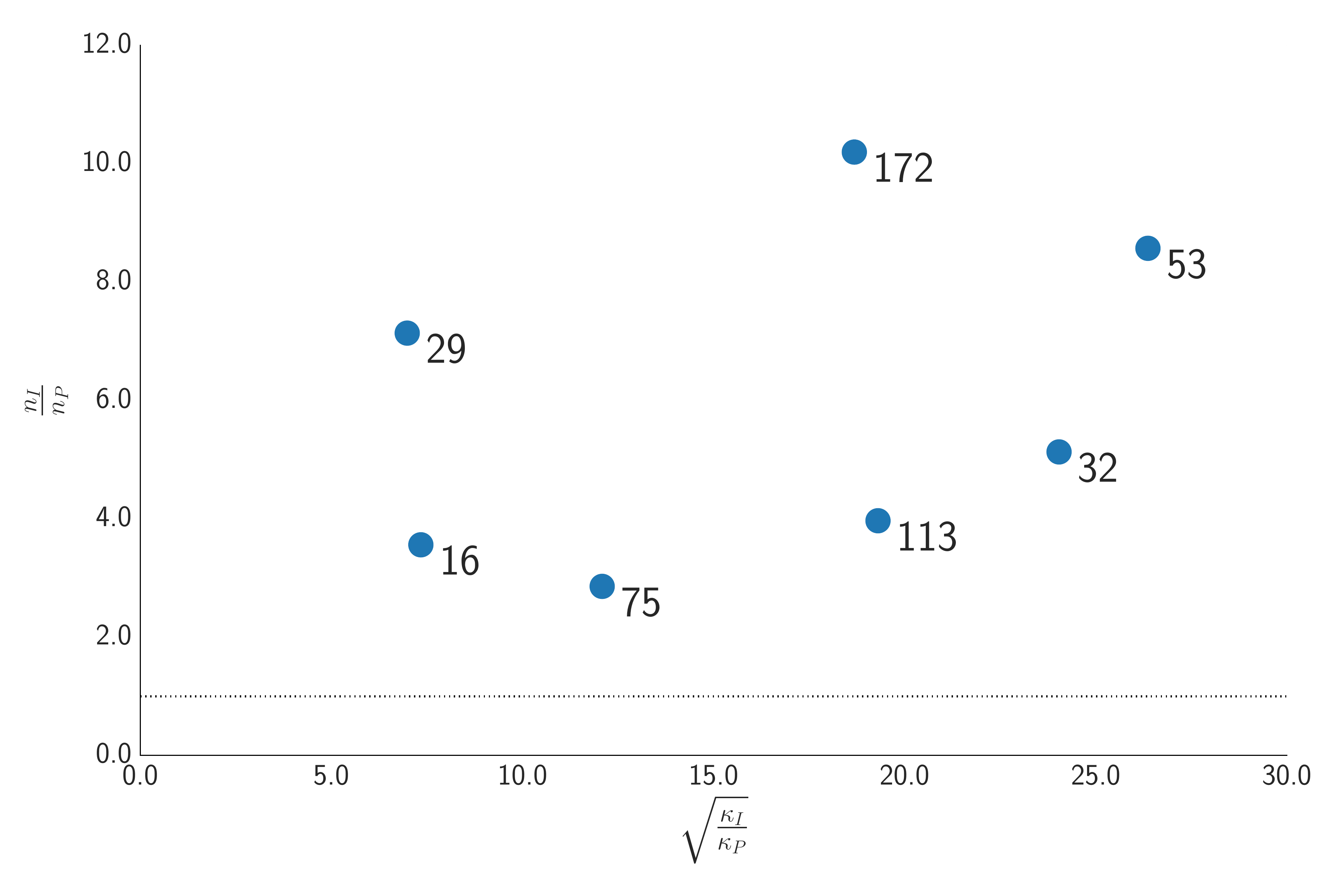}
        \caption{Correlation between the computational saving and the condition number ratio at potential energy minima of molecular systems in gas phase on PM6 potential.}
        \label{fig:mol_minim_cond}
\end{figure}

For PM6 only, to highlight the correlation between performance gain and ill-conditioning, we also computed the ratio between the condition numbers for the unpreconditioned and preconditioned Hessians, $\kappa_{I} / \kappa_{P}$, at the minima, where
\begin{equation}
        \label{eq:cond_P}
        \kappa_{P} = \frac{\lambda_{P}^{\max}}{\lambda_{P}^{\min}} = \frac{\max \limits_{u^T P u = 1} u^{T} H u}{\min \limits_{u^T P u =1} u^{T} \tilde{H} u}
\end{equation}
where $\tilde{H}$ is a modified Hessian where the zero eigenvalue due to symmetries are removed. In Figure~\ref{fig:mol_minim_cond} we observe that the computational saving is more strongly correlated to the condition number ratio than to the system size.

\begin{figure}
        \centering
        \includegraphics[width=1.0\columnwidth]{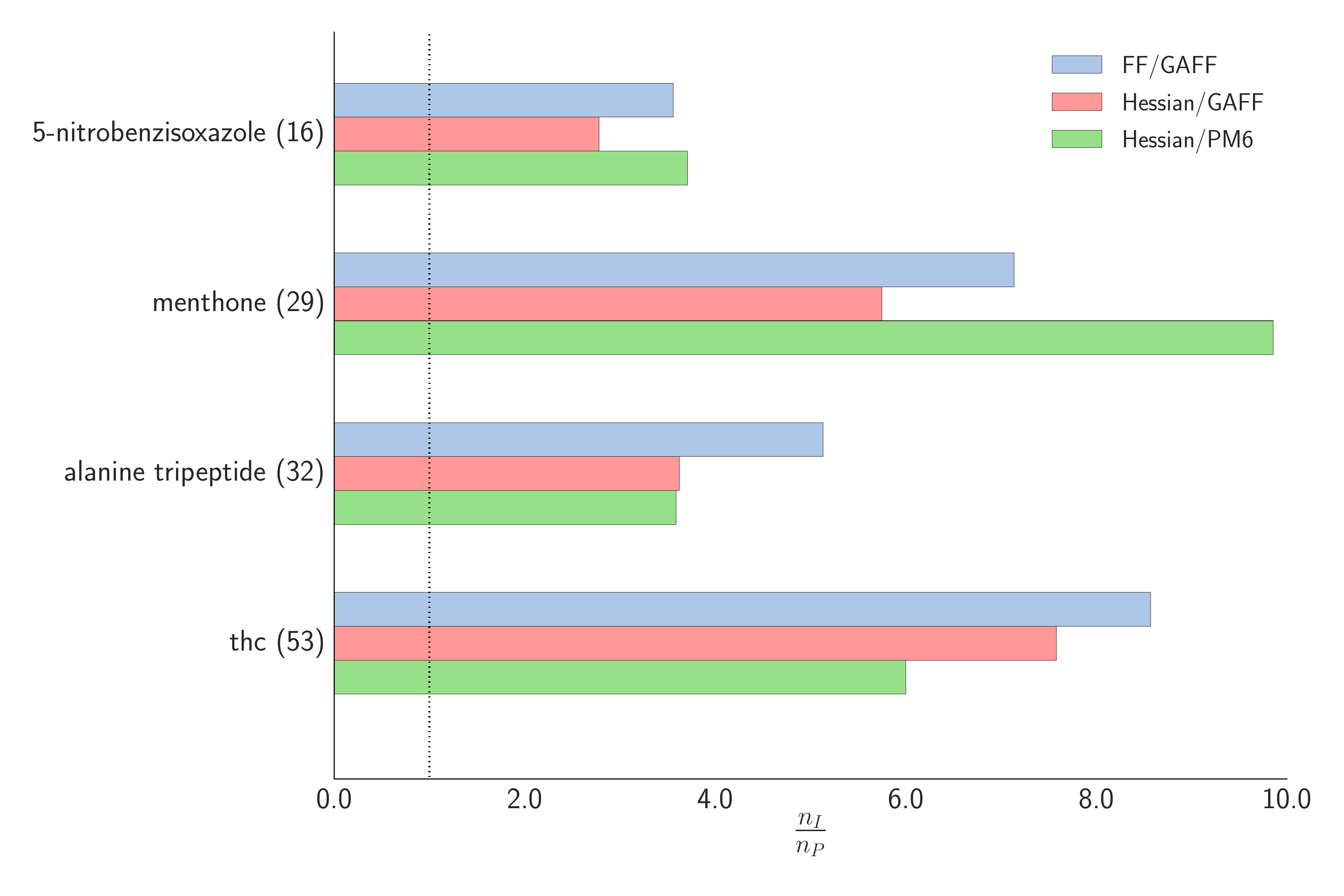}
        \caption{Computational gain of different optimisation algorithms compared to the unpreconditioned LBFGS method for geometry optimisation of molecular systems in gas phase on PM6 potential.}
        \label{fig:minim_mol_gas_lbfgs_newton}
\end{figure}

For the three smallest systems and the PM6 surface only we also compared our FF-based preconditioner against using the exact Hessian of the model potential (Hessian/GAFF) and a finite-difference hessian of PM6 (Hessian/PM6) as preconditioners for LBFGS method.
Zero eigenvalues of the Hessian matrices were shifted to a moderate positive number to avoid numerical instabilities. The results are collected in Table~\ref{tab:minim_mol_gas_lbfgs_newton} and Figure~\ref{fig:minim_mol_gas_lbfgs_newton}. Our FF-based clearly preconditioner outperforms both of these variants.

\begin{table}
\centering
\begin{tabular}{lllll}
        \toprule
        \multirow{2}{*}{System (\# of atoms)} & \multicolumn{4}{c}{LBFGS} \\
        & ID & FF/GAFF & Hessian/GAFF & Hessian/PM6 \\
        \midrule
        5-nitrobenzisoxazole (16) & 89 & 25 & 32 & 24 \\
        menthone (29) & 207 & 29 & 36 & 21 \\
        alanine tripeptide (32) & 395 & 77 & 109 & 110 \\
        thc (53) & 720 & 84 & 95 & 120 \\
        \bottomrule
        \caption{Comparison of total number of function/gradient calls of geometry optimisation of different optimisation algorithms for minimisation of small organic molecules on PM6 surface: unpreconditioned LBFGS (ID), FF-based preconditioned LBFGS (FF/GAFF), FF-Hessian based preconditioned LBFGS (Hessian/GAFF), PM6-Hessian based preconditioned LBFGS (PM6/GAFF). Convergence threshold was $|| \nabla E || = 10^{-4}$ eV \AA$^{-1}$ for all cases.}
        \label{tab:minim_mol_gas_lbfgs_newton}
\end{tabular}
\end{table}

\begin{figure}
        \centering
        \includegraphics[width=1.0\columnwidth]{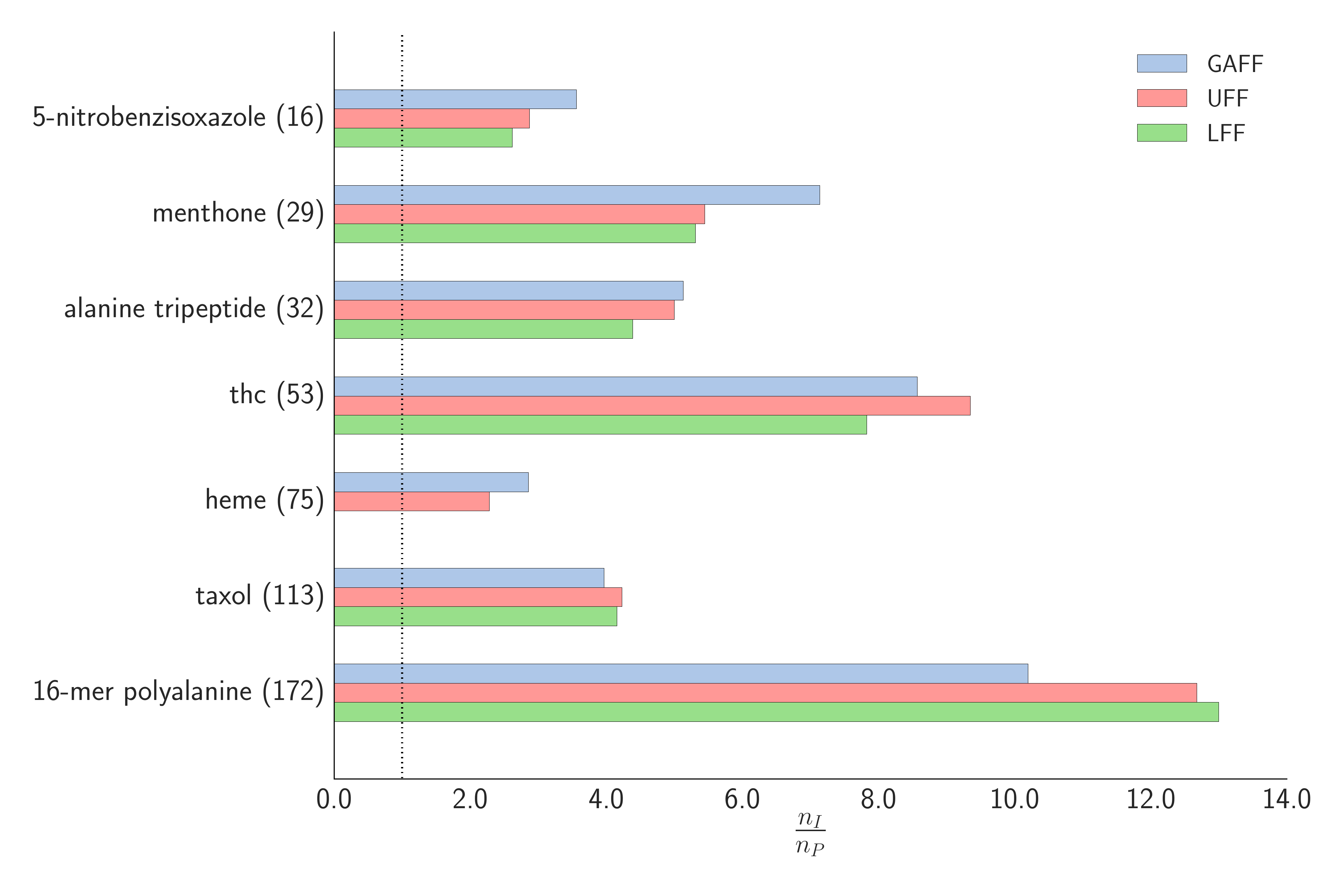}
        \caption{Computational gain of the FF-based preconditioner using different force fields for geometry optimisation of molecular systems in gas phase on PM6 potential.}
        \label{fig:minim_mol_gas_ffs}
\end{figure}

We also examined the effect of using different force fields from which to construct the FF-based preconditioner. Beside the GAFF force field, we investigated two general force fields: the universal force field (UFF)~\cite{Rappe92} and a general force field introduced by Lindh et al. (LFF)~\cite{Lindh95}. The results shown in Table~\ref{tab:minim_mol_gas_ffs} and Figure~\ref{fig:minim_mol_gas_ffs} indicate that there is no significant difference between the three force fields. We only mention that the LFF force-field includes all possible 2, 3 and 4--body interactions \cite{Lindh95}, resulting in a dense preconditioner matrix, which for larger systems and an efficient potential energy surface could become a performance bottleneck. By contrast, the preconditioners based on GAFF or UFF are sparse, hence their cost scales linearly with system size.

\begin{table}
\centering
\begin{tabular}{lllll}
        \toprule
        System (\# of atoms) & ID & FF/GAFF & FF/UFF & FF/LFF \\
        \midrule
        5-nitrobenzisoxazole (16) & 89 & 25 & 31 & 34 \\
        menthone (29) & 207 & 29 & 39 & 39 \\
        alanine tripeptide (32) & 395 & 77 & 79 & 90 \\
        thc (53) & 720 & 84 & 77 & 92 \\
        heme (75) & 500 & 175 & 219 & n.a. \\
        taxol (113) & 1662 & 419 & 393 & 400 \\
        16-mer polyalanine (172) & 3549 & 348 & 280 & 273 \\
        \bottomrule
        \caption{Comparison of the effect of different force field based preconditioners for the geometry optimisation of organic molecules in gas phase (total number function/gradient calls). Convergence threshold was $|| \nabla E || = 10^{-4}$ eV \AA$^{-1}$ for all cases.}
        \label{tab:minim_mol_gas_ffs}
\end{tabular}
\end{table}

\begin{figure}
        \centering
        \includegraphics[width=1.0\columnwidth]{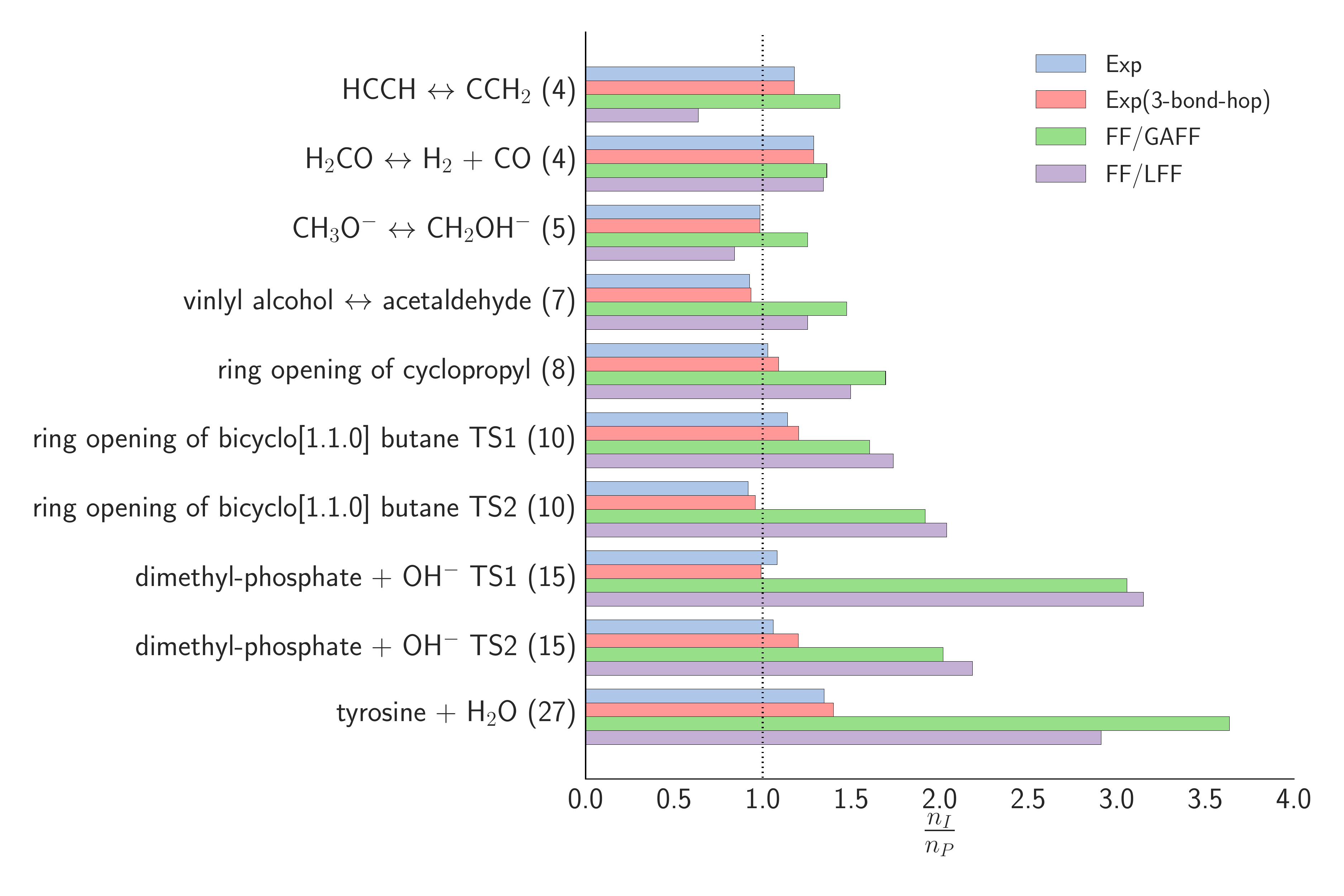}
        \caption{Computational gain of the Exp and FF-based preconditioners compared to the unpreconditioned superlinearly converging dimer method for molecular systems in gas phase on PM6 potential.}
        \label{fig:saddle_mol_gas_ffs}
\end{figure}

Finally, we tested how the different preconditioners perform when applied to transition state search, comparing again against ID (no preconditioning) and against the Exp preconditioner (with default parameter set and $\mu = 1$; unlike LBFGS, CG is invariant under rescaling of $\mu$).
The results are collected in Table~\ref{tab:saddle_mol_gas_ffs} and Figure~\ref{fig:saddle_mol_gas_ffs}. Overall the Exp preconditioner does not improve significantly over ID. We experimented with different parameters, e.g., adding connectivity information up to the 4-body interaction, but observed no improvements.
Both FF-based preconditioners are again comparable and yield a much improved convergence even for these relatively small systems. For instance, the gain is already 2--3-fold for dimethyl-phosphate and tyrosine hydrolyses.

\begin{table}
\centering
\begin{tabular}{lllll}
        \toprule
        System (\# of atoms) & ID & Exp & FF/GAFF & FF/LFF \\
        \midrule
        HCCH $\leftrightarrow$ CCH$_{2}$ (4) & 20 (79) & 17 (67) & 14 (55) & 32 (124) \\
        H$_{2}$CO $\leftrightarrow$ H$_{2}$ + CO (4) & 24 (94) & 19 (73) & 18 (69) & 18 (70) \\
        CH$_{3}$O$^{-}$ $\leftrightarrow$ CH$_{2}$OH$^{-}$ (5) & 18 (69) & 18 (70) & 14 (55) & 22 (82) \\
        vinyl alcohol $\leftrightarrow$ acetaldehyde (7) & 58 (227) & 62 (245) & 39 (154) & 46 (179) \\
        ring opening of cyclopropyl (8) & 52 (205) & 51 (199) & 31 (121) & 35 (137) \\
        ring opening of bicyclo[1.1.0] butane TS 1 (10) & 87 (332) & 77 (291) & 56 (207) & 52 (191) \\
        ring opening of bicyclo[1.1.0] butane TS 2 (10) & 66 (259) & 72 (282) & 35 (135) & 33 (127) \\
        dimethyl-phosphate + OH$^{-}$ TS 1 (15) & 395 (1541) & 361 (1425) & 127 (504) & 123 (489) \\
        dimethyl-phosphate + OH$^{-}$ TS 2 (15) & 355 (1379) & 329 (1302) & 172 (683) & 159 (631) \\
        tyrosine + H$_{2}$O (27) & 531 (2120) & 396 (1574) & 147 (583) & 185 (728) \\
        \bottomrule
        \caption{Number of steps of translations (and total number of function and gradient calls in parentheses) of saddle searches using different preconditioned variants of superlinearly converging dimer method. Convergence threshold was $|| \nabla E || = 10^{-4}$ eV \AA$^{-1}$ for all cases.}
        \label{tab:saddle_mol_gas_ffs}
\end{tabular}
\end{table}

\subsection{Molecular crystals}

\begin{figure}
        \centering
        \includegraphics[width=1.0\columnwidth]{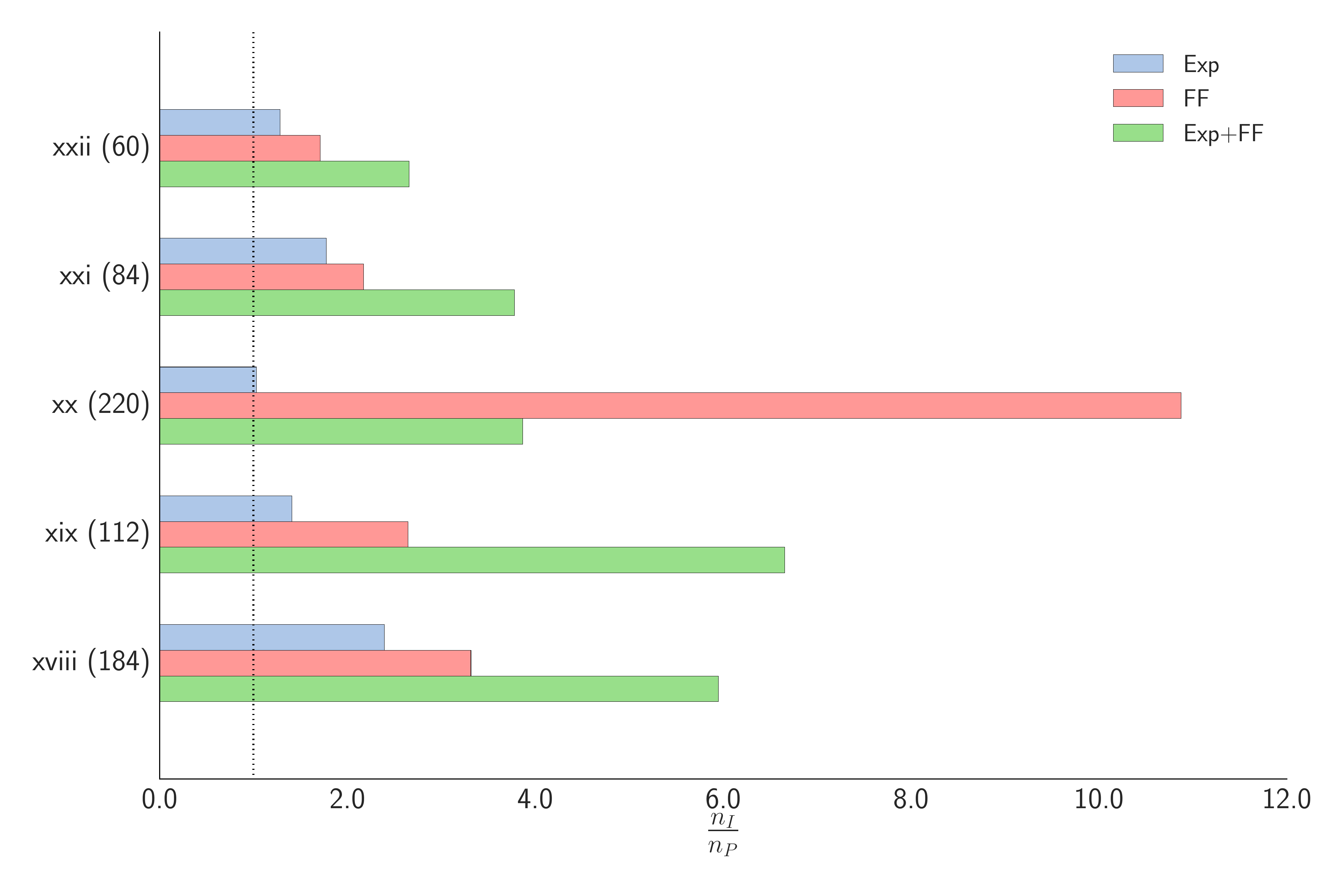}
        \caption{Computational gain of the Exp, FF and Exp+FF preconditioned over unpreconditioned LBFGS for geometry optimisation of molecular crystals on a DFT potential.}
        \label{fig:minim_mol_cryst_fix}
\end{figure}

We compared geometry optimisation with fixed unit cells using LBFGS, preconditioned with ID (unpreconditioned), Exp\cite{Packwood16} and FF(GAFF force field). For Exp the nearest neighbour distance ($r_{\mathrm{nn}}$) in Eq.~\ref{eq:exp_precond} was calculated from the initial structure, we specified $r_{\mathrm{cut}} = 2r_{\mathrm{nn}}$ and $A=3.0$. In addition, we also employed the Exp+FF preconditioner as defined in \eqref{eq:exp_ff_precond}. The results for different molecular crystals are shown in Table~\ref{tab:minim_mol_cryst_fix} and Figure~\ref{fig:minim_mol_cryst_fix}.
As expected, Exp reduces the number of optimisation steps compared to ID, although the improvement is significantly smaller for molecular crystals that for material systems\cite{Packwood16}. Interestingly, FF alone already leads to a significant speed-up over both ID and Exp even though the inter-molecular interaction is not captured well. This indicates that for molecular crystal optimisations preconditioning based on specific intramolecular information is crucial. The most successful method was the Exp+FF combination, which leads to a 3-7 fold speed up even for these relatively small test systems.

\begin{table}
\centering
\begin{tabular}{lllll}
        \toprule
        System (\# of atoms) & ID & Exp & FF/GAFF & Exp+FF/GAFF \\
        \midrule
        xxii (60) & 77 & 60 & 45 & 29 \\
        xxi (84) & 291 & 164 & 134 & 77 \\
        xx (220) &  174 & 169 & 16 & 45 \\
        xix (112) & 193 & 137 & 73 & 29 \\
        xviii (184) &  232 & 97 & 70 & 39 \\
        \bottomrule
        \caption{Total number of function/gradient calls of geometry optimisation of molecular crystals using different preconditioning strategies. Convergence criterion was $\| \nabla E \| = 1.0^{-3}$ eV \AA$^{-1}$.}
        \label{tab:minim_mol_cryst_fix}
\end{tabular}
\end{table}

\subsection{Material systems}
\label{sec:materials}

Finally, it is also interesting to investigate how an FF-based preconditioner compares against the Exp preconditioner for material systems, where Exp performs very well\cite{Packwood16}. We tested geometry optimisation of bulk silicon and a vacancy in bulk silicon with perturbed initial conditions, with increasing system size. The screened Tersoff potential was used as the potential energy, while the FF preconditioner was constructed from UFF. The results are shown in Table~\ref{tab:minim_material_silicon}. In both cases, the FF-based preconditioner yields a clear further speed-up over Exp for both systems.

\begin{table}
\centering
\begin{tabular}{lllllll}
        \toprule
        \multirow{2}{*}{System (\# of atoms)} & \multicolumn{3}{c}{Bulk geometry optimisation} & \multicolumn{3}{c}{Vacancy geometry optimisation} \\
        & ID & Exp & FF/UFF & ID & Exp & FF/UFF \\
        \midrule
        $2 \times 2 \times 2$ (64) & 32 & 17 & 10 & 34 & 15 & 9 \\
        $4 \times 4 \times 4$ (512) & 63 & 18 & 10 & 57 & 16 & 9 \\
        $8 \times 8 \times 8$ (4096) & 105 & 21 & 13 & 96 & 17 & 11 \\
        $16 \times 16 \times 16$ (32768) & 147 & 35 & 21 & 142 & 19 & 11 \\
        \bottomrule
        \caption{Total number of function/gradient calls geometry optimisation steps of bulk silicon and a bulk silicon vacancy using different preconditioning strategies. Convergence criterion was $\| \nabla E \| = 1.0^{-3}$ eV \AA$^{-1}$.}
        \label{tab:minim_material_silicon}
\end{tabular}
\end{table}

Another test system was bulk tungsten and a single interstitial site in bulk tungsten with perturbed initial conditions and different system sizes. The potential energy surface was provided by a GAP model that was trained on  DFT data. The preconditioner was based on a simple EAM potential:

\begin{equation}
        \label{eq:eam}
        V_{\mathrm{EAM}} = \sum_{i} \left[ \frac{1}{2} \sum_{j \neq i} \Phi(r_{ij}) + F \left( \sum_{j \neq i} \rho(r_{ij}) \right) \right]
\end{equation}
where $\Phi(r_{ij})$ is a pair potential, $F$ is the embedding function and $\rho({r_{ij}})$ is the electron charge density contribution from atom $j$ to atom $i$. Based on equation~\ref{eq:ff_precond} our FF-based preconditioner was defined as $P_{\mathrm{FF}} = \sum_{\alpha} \frac{\partial r_{\alpha}}{\partial \mathbf{r}} \otimes \frac{\partial r_{\alpha}}{\partial \mathbf{r}} \left| \frac{\partial^2 V_{\mathrm{EAM}}}{\partial r_{\alpha}^{2}} \right|$, where $\alpha$ runs over all $ij$ pairs. In the actual implementation $\Phi$, $F$ and $\rho$ functions are represented by splines so computing the corresponding curvature is fairly straightforward.

The results are presented in Table~\ref{tab:minim_material_tungsten}. For both the bulk and interstitial systems the number of function/gradient calls of the unpreconditioned optimisation increases with system size while the preconditioned optimisations require almost the same number of optimisation steps to achieve the same convergence criterion.

\begin{table}
\centering
\begin{tabular}{lllllll}
        \toprule
        \multirow{2}{*}{System (\# of atoms)} & \multicolumn{3}{c}{Bulk geometry optimisation} & \multicolumn{3}{c}{Interstitial geometry optimisation} \\
        & ID & Exp & FF/EAM & ID & Exp & FF/EAM \\
        \midrule
        $4 \times 4 \times 4$ (64) & 21 & 11 & 9 & 62 & 42 & 33 \\
        $8 \times 8 \times 8$ (512) & 32 & 12 & 8 & 72 & 43 & 36 \\
        $16 \times 16 \times 16$ (4096) & 56 & 12 & 9 & 116 & 46 & 33 \\
        \bottomrule
        \caption{Total number of function/gradient calls for geometry optimisation of bulk tungsten and interstitial defect in bulk tungsten using different preconditioning strategies. Convergence criterion was $\| \nabla E \| = 1.0^{-3}$ eV \AA$^{-1}$.}
        \label{tab:minim_material_tungsten}
\end{tabular}
\end{table}

\section{Conclusion}

We introduced a flexible preconditioner for molecular simulation based on empirical potentials that are widely implemented in popular molecular mechanical program packages. Our method, which can be considered a generalisation of Lindh et al.\cite{Lindh95}, decomposes the analytic Hessian of the empirical potential and modifies individual components to ensure their positivity. An advantage of this procedure is that it avoids the computation of second derivatives of the collective variables (or internal coordinates). The preconditioner yields significant improvements (at least 2 fold, and typically 5 fold decrease in function/gradient calls compared to unpreconditioned techniques), demonstrated thoroughly on a wide range of systems including molecules in gas phase, molecular crystals and materials, using different target potential energy surfaces (empirical, semiempirical and ab initio) as well as different optimisation tasks (geometry optimisations and saddle searches).

\section{Acknowledgement}

The authors thank Prof. Chris J Pickard for stimulating discussions. Computing facilities were provided by the Centre for Scientific Computing of the University of Warwick.

\clearpage

\bibliography{manuscript}

\clearpage

\end{document}